\documentclass[aps,prd,
onecolumn,
nofootinbib,amsmath,amssmyb,12pt]{revtex4-2}

\usepackage{color}
\usepackage{graphicx}
\usepackage{float}
\usepackage[usenames,dvipsnames]{xcolor}
\usepackage{multirow}
\usepackage{braket}
\usepackage{amsfonts}
\usepackage{mathtools}
\usepackage{stackengine}
\usepackage{feynmp-auto}
\usepackage{dcolumn}
\usepackage{bm}
\usepackage{tikz}
\usepackage{slashed}
\usepackage{hyperref}
\usetikzlibrary { decorations.pathmorphing, decorations.pathreplacing, decorations.shapes, }

\newcommand{\comma}{,\,}
\newcommand{\semicomma}{;\,}

\stackMath
\def\rldp{2.0ex}
\def\rlht{.8ex}
\def\rlwd{.8pt}
\setstackgap{L}{\rldp}
\def\uvbarR#1{%
  \def\stackalignment{r}\def\stacktype{S}\stackunder[-\rlwd]{%
    \def\stackalignment{c}\def\stacktype{L}\stackunder{#1}{\rule{\rlwd}{\rlht}}%
  }{\setbox0\hbox{$#1$}\rule{.5\wd0}{\rlwd}}%
}
\def\uvbarL#1{%
  \def\stackalignment{l}\def\stacktype{S}\stackunder[-\rlwd]{%
    \def\stackalignment{c}\def\stacktype{L}\stackunder{#1}{\rule{\rlwd}{\rlht}}%
  }{\setbox0\hbox{$#1$}\rule{.5\wd0}{\rlwd}}%
}
\def\uvbar#1#2{%
  \def\stacktype{S}\def\stackalignment{c}\def\useanchorwidth{T}\stackunder[0pt]{%
    \def\stacktype{L}\setbox0\hbox{${}#1{}$}%
    \stackunder[\rldp]{{}#1{}}{\rule{\wd0}{\rlwd}}%
  }{}
}
\newcommand\ubar[4]{\uvbarR{#1} \uvbar{#2}{#4} \uvbarL{#3}}

\begin{document}

\preprint{}

\title{Crossing Symmetry and Entanglement}
\author{Navin McGinnis}
 \email{nmcginnis@arizona.edu}
\affiliation{%
 Department of Physics, University of Arizona, Tucson, Arizona 85721, USA}%



\date{\today}

\begin{abstract}
We study the interplay between crossing symmetry and entanglement in $2 \to 2$ scattering within local quantum field theories that possess an $SU(N)$ global symmetry.  In particular, we recast scattering amplitudes of fixed helicity as quantum operations on the Hilbert space of internal quantum numbers, where the external states play the role of qudits. The entire space of $SU(N)$-invariant scattering operators between qudits is spanned by a minimal set of three quantum gates. Recoupling relations among quantum gates are shown to follow directly from the crossing properties of the underlying amplitudes and reveal that entanglement generated from separable states in one channel is necessarily intertwined with another.  Consequently, we argue any interacting quantum field theory that realizes an $SU(N)$ global symmetry must generate entanglement in at least one scattering channel.
\end{abstract}

\pacs{}
\keywords{}

\maketitle
\newpage

\section{Introduction}
Symmetry has long been one of the paramount mathematical tools in particle physics.  It provides a framework for organizing the spectrum of elementary particles, formulating the structure of the Standard Model (SM), and identifying the conservation laws that govern interactions. Further, the predictive power of symmetry, which has long guided the discovery of new phenomena, is unrivaled. Yet the origin of symmetry in Nature remains a mystery, and no physical principle is known that offers a hint.  Nevertheless, the search for such a principle has persisted throughout the history of science.

Recent work suggests that symmetries of particle and nuclear physics may arise from information-theoretic principles.  Growing evidence across a range of examples indicates a correlation between fundamental symmetries and the entanglement generated in the scattering of particles.  In particular, enhanced symmetries observed in nucleon scattering have been found to coincide with the preservation of separable states, hinting at an underlying entanglement-suppression principle~\cite{Beane:2018oxh,Low:2021ufv}.  Minimal entanglement has been further explored in low-energy QCD~\cite{Beane:2021zvo,Liu:2022grf,Bai:2022hfv,Bai:2023rkc,Bai:2023tey,Liu:2023bnr,Kirchner:2023dvg,Hu:2024hex,Hu:2025lua,Cavallin:2025kjn}, scalar field theories~\cite{Carena:2023vjc,Kowalska:2024kbs,Chang:2024wrx}, the SM quark sector~\cite{Thaler:2024anb}, string scattering~\cite{Bhat:2024agd}, and black holes~\cite{Aoude:2020mlg}.  In contrast, correlations between symmetry and maximal entanglement have been identified in the type-II two-Higgs-doublet model~\cite{Carena:2025wyh}, electroweak interactions~\cite{Cervera-Lierta:2017tdt,Miller:2023ujx}, Yang-Mills~\cite{Nunez:2025dch}, and in connection with electroweak phase transitions~\cite{Liu:2025pny}. In~\cite{McGinnis:2025brt}, we introduced a purely $S$-matrix formulation of these insights, connecting minimally entangling quantum gates to the emergence of $SU(N)$ global symmetries.

These proposals reflect the perspective of \textit{it from bit}, attributed to J. A. Wheeler, asserting that physical reality has its fundamental origin in information.  It is then natural to ask how the correlation between symmetry and entanglement is connected to the established principles of quantum field theory.  Among the general properties of the $S$-matrix (pioneered by Wheeler himself), crossing symmetry occupies a central place. Heuristically, crossing is tied to the notion that particles moving forward in time are indistinguishable from those of antiparticles moving backward.  More formally, it is seen as a consequence of locality and causality, relating scattering amplitudes describing distinct on-shell processes as analytic continuations of a single complex function. Indeed, Ref.~\cite{Chang:2024wrx} found indications that crossing symmetry, when combined with entanglement suppression, may impose strict constraints on the space of consistent theories. Similar comments can also be found in~\cite{Carena:2025wyh}. However, placing these observations on a rigorous footing in quantum field theory remains to be seen.

In this work, building on~\cite{McGinnis:2025brt} we develop a framework to describe crossing symmetry in the context of quantum information.  We consider two-particle scattering processes in which external states carry internal quantum numbers forming a finite-dimensional Hilbert space.  Treating these degrees of freedom as qudits allows the $S$-matrix to be represented as an operator on a tensor-product space, making its entangling properties explicit at the level of scattering amplitudes. The conventional decomposition of $SU(N)$-invariant amplitudes into irreducible representations realizes a natural interpretation in terms of quantum gates. Recoupling relations between quantum gates arise as a consequence of crossing symmetry, linking the analytic and algebraic properties of the $S$-matrix to quantum correlations.

A direct consequence is that crossing symmetry ties the presence or absence of entanglement across different scattering channels.  If one channel acts as an entanglement-suppressing operation, the crossed channel necessarily generates entanglement.  We find that only a trivial, non-interacting theory can preserve separability in all channels.  This establishes a correspondence between crossing symmetry and the existence of entanglement for interactions obeying a global symmetry.

This work seems timely in light of other efforts to frame fundamental aspects of the $S$-matrix through the lens of quantum information. This includes connections between cross sections and new area laws for scattering~\cite{Low:2024mrk,Low:2024hvn} and entanglement entropy~\cite{Kowalska:2025qmf},  positivity of scattering amplitudes~\cite{Aoude:2024xpx}, as well as extending the context of scattering to other quantum resources such as quantum magic~\cite{Busoni:2025dns,Liu:2025bgw,Liu:2025qfl,Gargalionis:2025iqs,Nunez:2025xds}. These efforts echo the need for new theoretical tools complementing the growing body of work applying quantum information as a new tool in collider physics~\cite{Barr:2024djo,Afik:2025ejh}. Finally, we note that crossing symmetry has recently been investigated from a purely quantum information-theoretic perspective~\cite{crossing_QI}.

The remainder of the paper is organized as follows.  Section~\ref{sec:2} reviews the Lehmann-Symanzik-Zimmerman (LSZ) formulation of crossing symmetry for scattering amplitudes with particular attention to global $SU(N)$ symmetries and derives the recoupling relations between invariant tensor bases.  Section~\ref{sec:3} reformulates two-particle scattering as a quantum operation on the Hilbert space of internal quantum numbers, identifying a minimal operator algebra governing entanglement generation.  Section~\ref{sec:4} combines these results to show that crossing symmetry and complete entanglement suppression cannot coexist except in the free theories. We conclude in section~\ref{sec:conclusions}.


\section{Crossing with global symmetries}
\label{sec:2}
Before drawing explicit connections to quantum information and entanglement, in this section we review aspects of crossing symmetry for 2-2 scattering amplitudes within the LSZ formalism. Although this serves mainly to setup notation and derive useful identities for later sections, we will give particular attention to crossing relations in the presence of global symmetries, an aspect that is often less emphasized.

\subsection{$S$-matrix for qudits}
Consider the scattering of relativistic particle states labeled by momenta, helicity, and an internal quantum number $\ket{p,\lambda;i}=\ket{p,\lambda}\otimes\ket{i}$. We will assume that all scattering states carry an internal label of the same dimension, $\text{dim}\{\ket{i}\}=N$, and refer to the finite dimensional Hilbert space spanned by these states as the \textit{flavor}- or, interchangably, \textit{qudit}-space. Assuming that the momentum-helicity components of the scattering states are normalized in a Lorentz-invariant way, and that qudits form a complete orthonormal set, the one-particle states are normalized in the conventional way
\begin{equation}    \braket{p^{\prime},\lambda^\prime;j|p,\lambda;i}=(2\pi)^{3}2E_{p}\delta^{(3)}(\vec{p}-\vec{p}^{\prime})\delta_{\lambda\lambda^{\prime}}\delta_{ij}.
    \label{eq:def_rel}
\end{equation}
Elements of the $S$-matrix between two particle states are then given by
\begin{flalign}\nonumber
    \bra{\{3\},\{4\}}S&\ket{\{1\},\{2\}}=(2\pi)^{4}\delta^{(4)}(p_{1}+p_{2}-p_{3}-p_{4})\bra{i_{3}i_{4}}\bra{p_{3},\lambda_{3};p_{4},\lambda_{4}}S\ket{p_{1},\lambda_{1};p_{2},\lambda_{2}}\ket{i_{1}i_{2}},
\end{flalign}
where the two-particle states are constructed from the direct product, e.g. $\ket{\{1\},\{2\}}=\ket{p_{1},\lambda_{1}; i_{1}}\otimes\ket{p_{2},\lambda_{2}; i_{2}}$. 

Our focus will be on $S$-matrix elements in the qudit space. Thus, to simplify the treatment we note that the kinematic part of the $S$-matrix, defined as the sub-matrix elements between momentum and helicity degrees of freedom, can be recast as a finite-dimensional operator on the subspace $H_{N}\otimes H_{N}\equiv\{\ket{i}\otimes\ket{j}\}$, whose coefficients are parameterized by the invariant-mass-squared of the two particle system, $s$, and a scattering angle~\footnote{See~\cite{Weinberg:1995mt} for an alternative argument based on the compactness of kinematic space.}
\begin{equation}
	S_{kl,ij}(s,\Omega) = \bra{kl}\textbf{S}(s,\Omega)\ket{ij},
\end{equation}
where $\textbf{S}(s,\Omega) = \bra{p_3,\lambda_{3};p_{4},\lambda_4}S\ket{p_{1},\lambda_{1};p_{2},\lambda_{2}}$~\cite{McGinnis:2025brt}. 

For further convenience we decompose $S$-matrix elements by subtracting off the disconnected contributions to the scattering process
\begin{equation}
    S-1=i\mathcal{T},
\end{equation}
defining the fully connected, interacting components by the operator $\mathcal{T}$. Thus, the main objects we will use to discuss crossing relations will be the 2-2 scattering amplitudes similarly defined as finite dimensional operators on the qudit space
\begin{equation}
    \bra{kl}i\mathbf{T}(s,\Omega)\ket{ij}=(2\pi)^{4}\delta^{(4)}(p_{1}+p_{2}-p_{3}-p_{4})i\mathcal{M}_{kl,ij}(s,\Omega),
\end{equation}
where $\textbf{T}(s,\Omega) = \bra{p_3,\lambda_{3};p_{4},\lambda_4}\mathcal{T}\ket{p_{1},\lambda_{1};p_{2},\lambda_{2}}$.

We note that the defining relations for the qudit states
\begin{equation}
	\braket{i|j} = \delta_{ij},\quad\quad\quad \sum_{i}\ket{i}\bra{i} = \mathbb{I}
\end{equation}
are chosen simply as normalization condition for the one-particle scattering states. Nevertheless, these relations are invariant under a global $SU(N)$ rotation, $\ket{i}^{\prime}=U\ket{j}$. Under this action, we define the qudit states to transform in the fundamental representation of $SU(N)$, $\ket{i}\sim \mathbf{N}$, while their complex conjugates transform in the anti-fundamental representation, $\ket{i}^*=\ket{\bar{i}}\sim \bar{\mathbf{N}}$. Although the invariance of the normalization condition does not imply a symmetry for the $S$-matrix, we will see that utilizing this property of the qudit states will provide a technical advantage when discussing various relations between scattering amplitudes.

\subsection{\label{recoup_cross}Crossing and recoupling relations}
Crossing relations for scattering amplitudes with respect to global symmetries will be most transparent in the context of scalar field theories. Let us temporarily narrow our focus to a theory of complex scalar fields, and defer the generalization to any helicity to a later section. The fundamental requirement for any quantum field theory we may consider is one whose fields (conjugates) create properly normalized one-particle (anti-particle) states
\begin{flalign}
	\bra{p,\bar{j}}\phi_{i}\ket{0} &= \delta_{ij}e^{-ipx}\\
	\bra{p,j}\phi^{\dagger}_{i}\ket{0} &= \delta_{ij}e^{-ipx}.
\end{flalign}
This requirement plus the invariance of the defining relations for qudit states automatically enforces that the fields (conjugates) transform as a fundamental (anti-fundamental) irrep under the action of $SU(N)$
\begin{flalign}\nonumber
	\phi^{\prime}_{i}&\to U_{ij}\phi_{j}\\
	\phi^{\dagger\prime}_{\bar{i}}&\to U^{\dagger}_{ji}\phi^{\dagger}_{\bar{j}}.
\end{flalign}

Consider the 2-2 scattering amplitude given by the process $\phi_{i}(p_{1}) + \phi_{j}(p_{2}) \to \phi_{k}(p_{3}) + \phi_{l}(p_{4})$, which is formally obtained from the LSZ reduction formula~\footnote{Our conventions follow those of~\cite{Srednicki:2007qs}.}
\begin{flalign}\nonumber
i\mathcal{M}_{\bar{k}\bar{l},ij} \propto i^4\int d\tilde{x} \prod_{\mathfrak{o}=3,4} e^{-ip_{\mathfrak{o}}x_{\mathfrak{o}}}(-\partial^{2}_{\mathfrak{o}} + m_{\mathfrak{o}}^{2})\prod_{\mathfrak{i}=1,2}&e^{ip_{\mathfrak{i}}x_{\mathfrak{i}}}(-\partial^{2}_{\mathfrak{i}} + m_{\mathfrak{i}}^{2})\\
&\times\bra{0}T\phi_{\bar{k}}^{\dagger}(x_{3})\phi_{\bar{l}}^{\dagger}(x_{4})\phi_{i}(x_{1})\phi_{j}(x_{2})\ket{0}_{f.c.},
\label{eq:particle_particle}
\end{flalign}
ignoring the factor of $(2\pi)^{4}\delta^{(4)}\left(\sum p_{\mathfrak{o}} - p_{\mathfrak{i}}\right)$ resulting from total momentum conservation. Although we are considering particle-particle scattering between qudit states in $H_{N}\otimes H_{N}$, we see that the scattering amplitude transforms as $\bar{\mathbf{N}}\otimes\bar{\mathbf{N}}\otimes \mathbf{N}\otimes \mathbf{N}$. Thus, under the action of $SU(N)^{4}$ we have that 
\begin{equation}
\mathcal{M}^{\prime}_{\bar{k}\bar{l},ij} \to (U_3^{\dagger})_{pk}(U_4^{\dagger})_{ql}(U_1)_{ir}(U_2)_{js}\mathcal{M}_{\bar{p}\bar{q},rs},
\label{eq:eq_class}
\end{equation}
reflecting the fact that scattering amplitudes are invariant under arbitrary field redefinitions. Without introducing any underlying lagrangian with a specified symmetry, Eq.~\ref{eq:eq_class} simply illustrates that the scattering processes represented on either side are defined with respect to different qudit bases of one-particle states. The LSZ formula holds provided that one-particle states are properly normalized and Eq.~\ref{eq:eq_class} reflects the transformations required for particle states to maintain the proper normalization in their respective bases.

Regardless of any symmetry in the underlying theory, it will be useful to decompose the scattering amplitude according the transformation properties of the fields. To do this, consider decomposing the pair of incoming fields into irreducible combinations
\begin{equation}
	\phi_{i}(x_{1})\phi_{j}(x_{2}) = \left[P_{\textbf{S}}+ P_{\textbf{A}}\right]_{ij,rs}\phi_{r}(x_{1})\phi_{s}(x_{2}) ,
\end{equation}
where
\begin{equation}
(P_{\textbf{S}})_{ij,rs} = \frac{\delta_{ir}\delta_{js} + \delta_{jr}\delta_{is}}{2},\quad\quad (P_{\textbf{A}})_{ij,rs} = \frac{\delta_{ir}\delta_{js} - \delta_{jr}\delta_{is}}{2},
\end{equation}
are the projection operators corresponding to the symmetric and antisymmetric invariant subspaces defined by $\mathbf{N}\otimes \mathbf{N}=\mathbf{S}\oplus\mathbf{A}$. Applying the analogous relation to the outgoing states ($\bar{\mathbf{N}}\otimes\bar{\mathbf{N}}=\mathbf{S}\oplus\mathbf{A}$), the 4-point function in Eq.~\ref{eq:particle_particle} can be expressed as

\begin{flalign}\nonumber
\bra{0}T\phi^{\dagger}_{\bar{k}}(x_{3})\phi^{\dagger}_{\bar{l}}(x_{4}) \phi_{i}(x_{1})&
\phi_{j}(x_{2})\ket{0}\\  &= \left[P_{\textbf{S}}+ P_{\textbf{A}}\right]_{\bar{k}\bar{l},\bar{p}\bar{q}}\left[P_{\textbf{S}}+ P_{\textbf{A}}\right]_{ij,rs}\bra{0}T\phi^{\dagger}_{\bar{p}}(x_{3})\phi^{\dagger}_{\bar{q}}(x_{4}) \phi_{r}(x_{1})\phi_{s}(x_{2})\ket{0},
\label{eq:sym_asym}
\end{flalign}
which clearly implies the same relation for the scattering amplitude
\begin{equation}
\mathcal{M}_{\bar{k}\bar{l},ij} = \left[P_{\textbf{S}}+ P_{\textbf{A}}\right]_{\bar{k}\bar{l},\bar{p}\bar{q}}\left[P_{\textbf{S}}+ P_{\textbf{A}}\right]_{ij,rs}\mathcal{M}_{\bar{p}\bar{q},rs}.
\end{equation}
The lack of invariance under a global $SU(N)$ is transparent from the presence of the off-diagonal projections $\sim P_{\textbf{S}}P_{\textbf{A}}$.
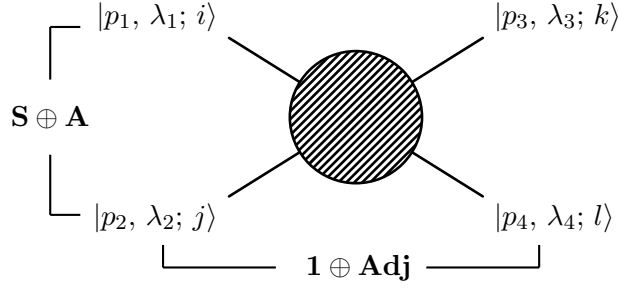
\begin{figure}[t]
    \centering
    \begin{tikzpicture}
	\node (diag) {
   	 \begin{fmffile}{blob_diagram}
 	 	\begin{fmfgraph*}(120,60)
   		 \fmfleft{i1,i2}   
    		\fmfright{o1,o2}  

    		\fmfv{decor.shape=circle, decor.filled=shaded, decor.size=50pt}{v} 

		    \fmf{plain}{i1,v}
		    \fmf{plain}{i2,v}
		    \fmf{plain}{v,o1}
		    \fmf{plain}{v,o2}

    		\fmfv{label=$\ket{p_{2}\comma\lambda_{2}\semicomma j}$, label.dist=5pt}{i1}
   		 \fmfv{label=$\ket{p_{1}\comma\lambda_{1}\semicomma i}$, label.dist=5pt}{i2}
		 \fmfv{label=$\ket{p_{4}\comma\lambda_{4}\semicomma l}$, label.dist=5pt}{o1}
	         \fmfv{label=$\ket{p_{3}\comma\lambda_{3}\semicomma k}$, label.dist=5pt}{o2}
  	\end{fmfgraph*}
	\end{fmffile}\vspace{0.25cm}
	};
	\draw[thick] (-4.0, 0.5)--(-4.0,1.2); 
	\draw[thick] (-4.0, -1.3)--(-4.0,-0.5); 
	
	\draw[thick] (-4.0, 1.2)--(-3.6,1.2); 
	\draw[thick] (-4.0, -1.3)--(-3.6,-1.3); 
	 \path (-4.0,0) node {$\textbf{S} \oplus \textbf{A}$};
	\draw[thick]  (-2.5,-2) -- (-1,-2);%
	\draw[thick]  (1,-2)  -- (2.5,-2);%
	
	\draw[thick]  (-2.5,-2)  -- (-2.5,-1.7);%
	\draw[thick]  (2.5,-2)  -- (2.5,-1.7);%
   	 \path (0.1,-2) node {$\textbf{1} \oplus \textbf{Adj}$};
	\end{tikzpicture}
	\hspace{2.0cm}
    \caption{Scattering of qudits $\ket{i,j}\to\ket{k,l}$ with different tensor product basis projections.}
    \label{fig:qudit_scattering}
\end{figure}

Note that pairing of incoming and outgoing states in the product $\bar{\mathbf{N}}\otimes\bar{\mathbf{N}}\otimes \mathbf{N}\otimes \mathbf{N}$ reflects only one choice of basis in the full decomposition. Another projection is given by pairing an incoming state together with an outgoing state. For example, pairing the indices $k\leftrightarrow i$ and $l\leftrightarrow j$ we obtain the decomposition
\begin{flalign}\nonumber
\bra{0}T\phi^{\dagger}_{\bar{k}}(x_{3})\phi^{\dagger}_{\bar{l}}(x_{4}) \phi_{i}(x_{1})&\phi_{j}(x_{2})\ket{0}\\  &= \left[P_{\textbf{1}}+ P_{\textbf{Adj}}\right]_{\bar{k}i,\bar{p}r}\left[P_{\textbf{1}}+ P_{\textbf{Adj}}\right]_{\bar{l}j,\bar{q}s}\bra{0}T\phi^{\dagger}_{\bar{p}}(x_{3})\phi^{\dagger}_{\bar{q}}(x_{4}) \phi_{r}(x_{1})\phi_{s}(x_{2})\ket{0},
\label{eq:sing_adj}
\end{flalign}
where
\begin{equation}
(P_{\textbf{1}})_{\bar{k}i,\bar{p}r} = \frac{\delta_{\bar{k}i}\delta_{\bar{p}r}}{N},\quad\quad (P_{\textbf{Adj}})_{ki,pr} = \delta_{\bar{k}p}\delta_{\bar{i}r} - \frac{\delta_{\bar{k}i}\delta_{\bar{p}r}}{N}
\end{equation}
are the projection operators corresponding to the singlet and adjoint invariant subspaces defined by $\bar{\mathbf{N}}\otimes \mathbf{N}=\mathbf{1}\oplus\mathbf{Adj}$. Of course a similar decomposition for the pairing $k\leftrightarrow j$ and $l\leftrightarrow i$. 

In general, the pairing does not conflict with the time ordering. As the previous example suggests, the decomposition 
\begin{equation}
	\phi^{\dagger}_{\bar{i}}\phi\ubar{^{\dagger}_{\bar{j}}}{\phi^{\dagger}_{\bar{l}}\cdots\phi_{p}\phi}{_q}{} \phi_{r}=\left[P_{\textbf{1}}+ P_{\textbf{Adj}}\right]_{\bar{j}q,\bar{s}t}	\phi^{\dagger}_{\bar{i}}\phi^{\dagger}_{\bar{s}}\phi^{\dagger}_{\bar{l}}\cdots\phi_{p}\phi_{t}\phi_{r}.
\end{equation}
simply projects onto a particular choice of basis in the total tensor product of of the internal quantum numbers. The choice of projection is arbitrary due to the fact that for finite dimensional vector spaces $V_{1}\otimes V_{2}\cdots \otimes V_{n} \simeq V_{\pi(1)}\otimes V_{\pi(2)}\cdots \otimes V_{\pi(n)}$, for any permutation $\pi$ on $\mathbb{Z}_{n}$. Since this choice is arbitrary, we arrive at the following relation
\begin{equation}
\left[P_{\textbf{S}}+ P_{\textbf{A}}\right]_{\bar{k}\bar{l},\bar{p}\bar{q}}\left[P_{\textbf{S}}+ P_{\textbf{A}}\right]_{ij,rs}\mathcal{M}_{\bar{p}\bar{q},rs} = \left[P_{\textbf{1}}+ P_{\textbf{Adj}}\right]_{\bar{k}i,\bar{p}r}\left[P_{\textbf{1}}+ P_{\textbf{Adj}}\right]_{\bar{l}j,\bar{q}s}\mathcal{M}_{\bar{p}\bar{q},rs}.
\end{equation}
Clearly, there must exist a change of basis between the two decompositions of the scattering amplitude implied by $\bar{\mathbf{N}}\otimes\bar{\mathbf{N}}\otimes \mathbf{N}\otimes \mathbf{N} = (\mathbf{S}\oplus\mathbf{A})\otimes(\mathbf{S}\oplus\mathbf{A}) = (\mathbf{1}\oplus\mathbf{Adj})\otimes(\mathbf{1}\oplus\mathbf{Adj})$. In general, the change of basis is known as a \textit{recoupling relation}.~\footnote{Recoupling relations for $SU(2)$ have been widely studied since Wigner's seminal work~\cite{Wigner:1936dx}, which introduced the change of basis coefficients now referred to as the Wigner 3-j and 6-j symbols.} In Appendix~\ref{sec:appendix_A}, we derive the recoupling relations between the $(\textbf{S},\textbf{A})$ and $(\textbf{1},\textbf{Adj})$ bases, which we repeat here in more compact notation
\begin{equation}
	\begin{pmatrix}P_{\mathbf{1}} \\P_{\mathbf{Adj}} \end{pmatrix} = \mathbf{C}\cdot \begin{pmatrix}P_{\mathbf{S}} \\P_{\mathbf{A}} \end{pmatrix},
\end{equation}
where 
\begin{equation}
\renewcommand\arraystretch{1.8}
	\mathbf{C}=\begin{pmatrix}\frac{1}{N} &\frac{1}{N}\\ \frac{N-1}{N} & -\frac{N+1}{N} \end{pmatrix}.
\label{eq:recoup}
\end{equation}
Similarly
\begin{equation}
	\begin{pmatrix}P_{\mathbf{S}} \\P_{\mathbf{A}} \end{pmatrix} = \mathbf{C}^{\prime}\cdot \begin{pmatrix}P_{\mathbf{1}} \\P_{\mathbf{Adj}} \end{pmatrix},
\end{equation}
with
\begin{equation}
\renewcommand\arraystretch{1.8}
	\mathbf{C}^{\prime}=\begin{pmatrix}\frac{N+1}{2} &\frac{1}{2}\\ \frac{N-1}{2} & -\frac{1}{2} \end{pmatrix}
	\label{eq:recoup_prime}
\end{equation}

Before discussing the crossed channels, let us consider a simple example to see how these tools play out in practice. Consider a theory of $N$ complex scalar fields with the lagrangian
\begin{equation}
\mathcal{L} = -\partial_{\mu}\phi_{i}^{\dagger}\partial^{\mu}\phi_{i} -m^{2}\phi^{\dagger}_{i}\phi_{i} - \frac{\lambda}{4}(\phi^{\dagger}_{i}\phi_{i})^{2}
\end{equation}
The leading-order amplitude for the process at hand is
\begin{equation}
i\mathcal{M}_{\bar{k}\bar{l},ij} = -i\frac{\lambda}{2}\left[P_{\textbf{S}}+ P_{\textbf{A}}\right]_{kl,pq}\left[P_{\textbf{S}}+ P_{\textbf{A}}\right]_{ij,rs}\left(\delta_{pr}\delta_{qs} + \delta_{ps}\delta_{qr}\right)
\end{equation}
which reduces to
\begin{equation}
\mathcal{M}_{\bar{k}\bar{l},ij} = -\lambda(P_{\textbf{S}})_{\bar{k}l\bar{,}ij},
\label{eq:scalar_1}
\end{equation}
In contrast, applying the expansion in Eq.~\ref{eq:sing_adj} we obtain
\begin{equation}
	\mathcal{M}_{\bar{k}\bar{l},ij} = -\frac{\lambda}{2}\left[(N+1)(P_{\textbf{1}})_{\bar{k}i,\bar{l}j} + (P_{\textbf{Adj}})_{\bar{k}i,\bar{l}j}\right].
\label{eq:scalar_2}
\end{equation}
A couple comments are in order. We see that supplying an invariant theory to the 4-point functions indeed results in amplitudes which act diagonally in the decomposition of irreducible representations. Further, starting from Eq.~\ref{eq:scalar_1} and applying the recoupling relation given by Eq.~\ref{eq:recoup_prime} we recover Eq.~\ref{eq:scalar_2}.

In general, any 2-2 scattering channel invariant under an $SU(N)$ global symmetry can be decomposed in two ways as~\footnote{A statement of this fact without proof can be found in~\cite{Mandula:1981xt}.}
\begin{equation}
	\mathcal{M} = \mathcal{M}_{\textbf{S}}P_{\textbf{S}} + \mathcal{M}_{\textbf{A}}P_{\textbf{A}}
\end{equation}
or
\begin{equation}
	\mathcal{M} = \mathcal{M}_{\textbf{1}}P_{\textbf{1}} + \mathcal{M}_{\textbf{Adj}}P_{\textbf{Adj}},
\end{equation}
depending on the choice of pairing between fields in the underlying 4-point function. The scalar amplitudes in these relations have been defined by
\begin{equation}
\mathcal{M}_{R} = \bra{kl}\mathcal{M}\ket{ij}(P_{R})_{kl,ij}/\text{Tr}(P_{R}),\quad\text{or}\;\; \mathcal{M}_{R} = \bra{kl}\mathcal{M}\ket{ij}(P_{R})_{ki,lj}/\text{Tr}(P_{R}),
\end{equation}
where $R = \mathbf{1},\mathbf{Adj}, \mathbf{S},$ or $\mathbf{A}$ depending on the specific pairing of representations. The recoupling relations between the bases of projection operators imposes the relations
\begin{flalign}
	\mathcal{M}_{\textbf{1}} &= \frac{(N+1)\mathcal{M}_{\mathbf{S}} + (N-1)\mathcal{M}_{\mathbf{A}}}{2}\\
	\mathcal{M}_{\textbf{Adj}} &= \frac{1}{2}\left(\mathcal{M}_{\mathbf{S}} - \mathcal{M}_{\mathbf{A}}\right).
\end{flalign}

Now, consider the crossed channel given by the scattering of a fundamental and anti-fundamental particles $\phi_{i}(p_{1}) + \phi^{\dagger}_{\bar{j}}(p_{2})\to\phi^{\dagger}_{\bar{k}}(p_{3}) + \phi_{l}(p_{4})$. The scattering amplitude obtained from the LSZ formula in this case is given by
\begin{flalign}\nonumber
i\mathcal{M}_{k\bar{l},i\bar{j}} \propto i^4\int d\tilde{x} \prod_{\mathfrak{o}=3,4} e^{-ip_{\mathfrak{o}}x_{\mathfrak{o}}}(-\partial^{2}_{\mathfrak{o}} + m_{\mathfrak{o}}^{2})&\prod_{\mathfrak{i}=1,2}e^{ip_{\mathfrak{i}}x_{\mathfrak{i}}}(-\partial^{2}_{\mathfrak{i}} + m_{\mathfrak{i}}^{2})\\
&\times\bra{0}T\phi_{k}(x_{3})\phi^{\dagger}_{\bar{l}}(x_{4})\phi_{i}(x_{1})\phi^{\dagger}_{\bar{j}}(x_{2})\ket{0}_{f.c.}.
\end{flalign}
The decomposition into projection operators follows along similar lines with the exception that pairing incoming states and outgoing states among themselves now results in $\bar{\mathbf{N}}\otimes\mathbf{N}=\mathbf{1}\oplus\mathbf{Adj}$, and vice versa when pairing an incoming state together with an outgoing state.

Finally, let us turn to the implications of crossing symmetry for scalar fields. Consider the two scattering channels
\begin{flalign}\nonumber
	\mathcal{M}^{(s)}:\quad\phi_{i}(p_{1}) + \phi_{j}(p_{2}) \to& \phi_{k}(p_{3}) + \phi_{l}(p_{4})\quad\quad s\text{-channel}\\
	\mathcal{M}^{(t)}:\quad	\phi_{i}(p_{1}) + \phi^{\dagger}_{\bar{k}}(p_{3}) \to& \phi^{\dagger}_{\bar{j}}(p_{2}) + \phi_{l}(p_{4})\quad\quad t\text{-channel},
\end{flalign}
where we have introduced the labeling via Mandelstam variables in the $s$-channel defined by
\begin{equation}
s=-(p_{1}+p_{2})^{2}, \quad\quad t = - (p_{1}-p_{3})^{2},
\end{equation}
and for the $t$-channel
\begin{equation}
t=-(p_{1}+p_{2})^{2}, \quad\quad s = - (p_{1}-p_{3})^{2}.
\end{equation}
Crossing symmetry imposes the constraint
\begin{equation}
	\mathcal{M}^{(s)}_{\bar{k}\bar{l},ij}\left(s=-(p_{1} + p_{2})^{2},t = -(p_{1} - p_{3})^{2}\right) = \mathcal{M}^{(t)}_{j\bar{l},i\bar{k}}\left(t=-(p_{1} + p_{2})^{2},s= -(p_{1} - p_{3})^{2}\right),
	\label{eq:crossing_scalars}
\end{equation}
which follows from analytic properties of scattering amplitudes in local quantum field theory.  Briefly, both amplitudes are obtained from time-ordered four-point functions via the LSZ reduction formula, differing only by whether the external fields is conjugated. The amplitudes are then different manifestations a single analytic Green’s function in complex momentum space. The two channels are related by analytically continuing the external momenta of the crossed particles.  In particular, starting from Eq.~\ref{eq:particle_particle}, we may perform the continuation $p_{2}\to -p_{2}$ and $p_{3}\to -p_{3}$, leading to $e^{ip_{2}x_{2}}\to e^{-ip_{2}x_{2}}$ and $e^{-ip_{3}x_{3}}\to e^{ip_{3}x_{3}}$, respectively.~\footnote{The analytic continuation $p_2 \to -p_2$ and $p_3 \to -p_3$ is understood within the connected domains of analyticity in complex momentum space guaranteed by the edge-of-the-wedge theorem. These domains allow continuation between the physical $s$- and $t$-channel regions while preserving the on-shell conditions within their common envelope of holomorphy. See~\cite{Bros:1965,Bros:1964iho} and~\cite{Mizera:2021ujs,Caron-Huot:2023ikn} for details.}  Analytically continuing the time-components consistent with the on-shell condition moves the amplitude between the physical regions; the result is the amplitude in which the original incoming (outgoing) particle 2 (3) is interpreted as an outgoing (incoming) conjugate particle.

We note that the rigorous proof of crossing for 2-2 scattering amplitudes, originally established by Bros, Epstein, and Glaser~\cite{Bros:1965,Bros:1964iho}, assumes that the underlying theory has a mass gap. We therefore take Eq.~\ref{eq:crossing_scalars} as a direct consequence of this general result along with its assumptions. In the present case, apart from analytic continuation we note that fully establishing crossing also requires use of the recoupling relations. For instance, if we choose to decompose the amplitudes consistently by pairing incoming states and outgoing states, we arrive at a familiar relation
\begin{equation}
\left[P_{\textbf{S}}+ P_{\textbf{A}}\right]_{kl,pq}\left[P_{\textbf{S}}+ P_{\textbf{A}}\right]_{ij,rs}\mathcal{M}^{(s)}_{\bar{p}\bar{q},rs} = \left[P_{\textbf{1}}+ P_{\textbf{Adj}}\right]_{jl,qs}\left[P_{\textbf{1}}+ P_{\textbf{Adj}}\right]_{ik,pr}\mathcal{M}^{(t)}_{q\bar{s},p\bar{r}}.
\end{equation}
where the analytic continuation of the momenta are implied in the labeling $\mathcal{M}^{(s/t)}$. Of course, a different choice of pairing on either side of Eq.~\ref{eq:crossing_scalars} would remove this ambiguity. However, for crossing to apply for any decomposition the recoupling relations between $(\textbf{S},\textbf{A})$ and $(\textbf{1},\textbf{Adj})$ bases must be applied. In particular, for amplitudes obeying an $SU(N)$ global symmetry, the recoupling relations between analytically continued scalar amplitudes take on a simple form
\begin{flalign}
	\mathcal{M}^{(s)}_{R_{s}}=&\sum_{R_t}C^{s,t}_{R_{s},R_{t}}\mathcal{M}^{(t)}_{R_{t}},
\end{flalign}
where the coefficents $C^{s,t}_{R_{s},R_{t}}$ are given by Eq.~\ref{eq:recoup} or \ref{eq:recoup_prime} depending on the processes at hand.
For instance, for the present case we have chosen to decompose the channels according to in-in and out-out states, and $R_{t}=\textbf{1},\textbf{Adj}$, $R_{s}=\textbf{S},\textbf{A}$. 
\subsection{Including helicity}
So far, for simplicity, we have restricted the discussion of crossing relations between scattering amplitudes involving scalar particles. The inclusion of helicity follows in close analogy.  The key difference is that the external wavefunctions now carry spinor or vector indices, so the analytic continuation must also account for the transformation of polarization states under Lorentz rotations.  As shown long ago by Trueman and Wick~\cite{Trueman:1964zzb}, and Hara~\cite{Hara:1964zza,Hara:1970gc} crossing a particle of spin $s_i$ to the opposite side of the amplitude introduces a phase determined by a corresponding Wigner rotation.

As an example to see this in action, we can consider particle-particle scattering of Dirac fermions with a global quantum number
$\bar{\psi}_{i}(p_{1}) + \bar{\psi}_{j}(p_{2}) \to \bar{\psi}_{k}(p_{3}) + \bar{\psi}_{l}(p_{4})$~\footnote{We have chosen $\bar{\psi}_i$ to create properly normalized states in the fundamental representation as a minor annoyance.}
\begin{flalign}\nonumber
i\mathcal{M}_{\bar{k}\bar{l},ij} \propto i^4\int d\tilde{x} \prod_{\mathfrak{o}=3,4}& e^{-ip_{\mathfrak{o}}x_{\mathfrak{o}}}\bar{u}_{\lambda_{\mathfrak{o}}}(p_{\mathfrak{o}})(-i\slashed{\partial}_{\mathfrak{o}} + m_{\mathfrak{o}})\\
&\times\bra{0}T\psi_{\bar{k}}(x_{3})\psi_{\bar{l}}(x_{4})\bar{\psi}_{i}(x_{1})\bar{\psi}_{j}(x_{2})\ket{0}_{f.c.}\prod_{\mathfrak{i}=1,2}e^{ip_{\mathfrak{i}}x_{\mathfrak{i}}}u_{\lambda_{\mathfrak{i}}}(p_{\mathfrak{i}})( i  \overleftarrow{\slashed{\partial}}_{\mathfrak{i}} + m_{\mathfrak{i}}),
\label{eq:fermion_particle_particle}
\end{flalign}
where we have implicitly chosen the spin axis of one-particle states to point along their direction of motion.
Crossing of particles $2\leftrightarrow 3$, can be heuristically obtained by the replacement $p_{2}\to-p_{2}$ and $p_{3}\to-p_{3}$ resulting in~\cite{Hara:1964zza,Hara:1970gc} 
\begin{flalign}
	e^{ip_{2}x_{2}}u_{\lambda_{2}}(p_{2})&\to e^{-ip_{2}x_{2}}e^{i\pi\lambda_{2}}v_{\lambda_{2}}(p_{2})\\
	e^{-ip_{3}x_{3}}\bar{u}_{\lambda_{3}}(p_{3})&\to e^{ip_{3}x_{3}}e^{-i\pi\lambda_{3}}\bar{v}_{\lambda_{3}}(p_{3}).
\end{flalign}
With these replacements, Eq.~\ref{eq:fermion_particle_particle} is clearly related to the amplitude for $\bar{\psi}_{i}(p_{1}) + \psi_{\bar{k}}(p_{3}) \to \psi_{\bar{j}}(p_{2}) + \bar{\psi}_{l}(p_{4})$
\begin{flalign}\nonumber
i\mathcal{M}_{j\bar{l},i\bar{k}} \propto i^4\int d\tilde{x} & \;e^{-ip_{4}x_{4}}\bar{u}_{\lambda_{4}}(p_{4})(-i\slashed{\partial}_{4} + m_{4})e^{-ip_{2}x_{2}}v_{\lambda_{2}}(p_{2})(-i\slashed{\partial}_{2} + m_{2})\\\nonumber
&\quad\quad\times\bra{0}T\bar{\psi}_{j}(x_{2})\psi_{\bar{l}}(x_{4})\bar{\psi}_{i}(x_{1})\psi_{\bar{k}}(x_{3})\ket{0}_{f.c.}\\
&\quad\quad\quad\quad\quad\times e^{ip_{1}x_{1}}u_{\lambda_{1}}(p_{1})( i  \overleftarrow{\slashed{\partial}}_{1} + m_{\mathfrak{1}})e^{ip_{3}x_{3}}\bar{v}_{\lambda_{3}}(p_{3})( i  \overleftarrow{\slashed{\partial}}_{3} + m_{3}).
\label{eq:fermion_particle_aparticle}
\end{flalign}
The relation of the two amplitudes is given by a complex Lorentz transformation~\cite{Trueman:1964zzb,Hara:1964zza,Hara:1970gc}, requiring an analytic continuation of Mandelstam variables via the time components of the 4-momenta.

To summarize the results concisely in the present context, for any fixed helicity amplitude cast as an operator on the qudit space
\begin{equation}
\mathcal{M}_{\bar{k}\bar{l},ij} = \bra{\bar{k}\bar{l}}\bra{\lambda_{3}\lambda_{4}}\mathcal{M}(s,\Omega)\ket{\lambda_{1}\lambda_{2}}\ket{ij}
\end{equation}
crossing relations among helicity components following the steps just outlined are given by
\begin{equation}
	\bra{\lambda_{3}\lambda_{4}}\mathcal{M}^{(s)}(s,\Omega)\ket{\lambda_{1}\lambda_{2}} = \epsilon \cdot e^{i\pi(\lambda_{1}+\lambda_{3})} \sum_{\{\lambda^{\prime}_{i}\}} e^{i\pi(\lambda^{\prime}_{1} + \lambda^{\prime}_{2})} \bra{\lambda^\prime_{2}\lambda^\prime_{4}}\mathcal{M}^{(t)}(t,\Omega)\ket{\lambda^{\prime}_{1}\lambda^{\prime}_{3}}\prod_{i} d^{s_{i}}_{\lambda^{\prime}_{i}\lambda_{i}}(\chi_{i}),
\end{equation}
where $\epsilon$ is an overall sign that depends on the specific helicities of a given process~\cite{Hara:1964zza,Hara:1970gc}, the rotation angles $\chi_{i}$ depend on the specific kinematics and masses for a given process, and $d^{s_{i}}_{\lambda^{\prime}_{i}\lambda_{i}}(\chi_{i})$ are the Wigner-D matrices resulting from reexpressing the $s$-channel helicity basis to the $t$-channel helicity basis~\cite{Trueman:1964zzb,Hara:1964zza,Burkhardt:1968,Hara:1970gc}~\footnote{Note, in applying the crossing relations to helicity amplitudes we have used the conventions in ~\cite{Hara:1964zza,Hara:1970gc} which do not include an extra phase factor for "particle 2" as in~\cite{Trueman:1964zzb}. See also~\cite{Bellazzini:2016xrt,deRham:2017zjm,Trott:2020ebl} for detailed explanations of different conventions.}.
Including the recoupling relations for the qudit quantum numbers, the full crossing relations for $SU(N)$ invariant amplitudes become
\begin{flalign}\nonumber
	\bra{kl}&\bra{\lambda_{3}\lambda_{4}}\mathcal{M}^{(s)}(s,\Omega)\ket{\lambda_{1}\lambda_{2}}\ket{ij}(P_{R_{s}})_{kl,ij}/\text{Tr}(P_{R_{s}}) =\epsilon \cdot e^{i\pi(\lambda_{1}+\lambda_{3})} \\&
	\times\sum_{R_{t},\{\lambda^{\prime}_{i}\}} e^{i\pi(\lambda^{\prime}_{1} + \lambda^{\prime}_{2})} C^{s,t}_{R_{s},R_{t}} \bra{pr}\bra{\lambda^\prime_{3}\lambda^\prime_{1}}\mathcal{M}^{(t)}(t,\Omega)\ket{\lambda^\prime_{4}\lambda^\prime_{2}}\ket{qs}(P_{R_{t}})_{pq,qrs}/\text{Tr}(P_{R_{t}})\prod_{i} d^{s_{i}}_{\lambda^{\prime}_{i}\lambda_{i}}(\chi_{i}).
\end{flalign}
Due to the tensor product structure between the momentum-helicity and qudit components, we may rewrite this as
\begin{flalign}\nonumber
 \mathcal{M}^{s}_{R_{s}}(s,\Omega&) = \sum_{R_{t}}C^{s,t}_{R_{s},R_{t}}(P_{R_{t}})_{pq,qrs}\bra{pr}\\&\left(\epsilon \cdot e^{i\pi(\lambda_{1}+\lambda_{3})}\sum_{\{\lambda^{\prime}_{i}\}}e^{i\pi(\lambda^{\prime}_{1} + \lambda^{\prime}_{2})} \bra{\lambda^\prime_{3}\lambda^\prime_{1}}\mathcal{M}^{(t)}(t,\Omega)\ket{\lambda^\prime_{4}\lambda^\prime_{2}}/\text{Tr}(P_{R_{t}}) \prod_{i} d^{s_{i}}_{\lambda^{\prime}_{i}\lambda_{i}}(\chi_{i})\right)\ket{qs}.
\end{flalign}
Further
\begin{equation}
	\mathcal{M}^{s}_{R_{s}}(s,\Omega) = \sum_{R_{t}}C^{s,t}_{R_{s},R_{t}} \tilde{\mathcal{M}}^{{t}}_{R_{t}}(t,\Omega),
	\label{eq:crossed}
\end{equation}
where we have defined the crossed-recoupled amplitude
\begin{flalign}\nonumber
	\tilde{\mathcal{M}}^{{t}}_{R_{t}}(t,\Omega) =& \bra{pr}\epsilon \cdot e^{i\pi(\lambda_{1}+\lambda_{3})}\\
		&\times\sum_{\{\lambda^{\prime}_{i}\}}e^{i\pi(\lambda^{\prime}_{1} + \lambda^{\prime}_{2})} \bra{\lambda^\prime_{3}\lambda^\prime_{1}}\mathcal{M}^{(t)}(t,\Omega)\ket{\lambda^\prime_{4}\lambda^\prime_{2}} \prod_{i} d^{s_{i}}_{\lambda^{\prime}_{i}\lambda_{i}}(\chi_{i})\ket{qs}(P_{R_{t}})_{pq,qrs}/\text{Tr}(P_{R_{t}}).
	\label{eq:crossed_recoupled}
\end{flalign}
Thus, although including helicities results in a number of technical details, the full crossing relations including the qudit quantum numbers can be compactly expressed in Eqs.~\ref{eq:crossed}~\&~\ref{eq:crossed_recoupled}. In the following sections, we will use these compact forms to draw connections between crossing symmetry and entanglement among the qudit quantum numbers.

\section{2-2 scattering channels as quantum operations}
\label{sec:3}
%
In this section, we review and expand on the structure of scattering amplitudes as quantum operations introduced in~\cite{McGinnis:2025brt}. Consider any fixed-helicity amplitude, $\mathcal{M}$, parameterized as an operator acting on $H_{N}\otimes H_{N}$. In order to understand the information-theoretic properties related to the explicit action of the amplitude on the qudit Hilbert space we may decompose it in terms of an operator basis. A natural choice would of course be the $N^{2}\times N^{2}$ operators $\ket{i}\bra{j}\otimes \ket{k}\bra{l}$. However, we will find it more convenient to decompose the amplitude in terms of the basis
\begin{equation}
\mathcal{M} = \tilde{\mathcal{M}}\mathbb{I}_{N}\otimes\mathbb{I}_{N}+\mathcal{A}_{a}T^{a}\otimes\mathbb{I}_{N} + \mathcal{B}_{a}\mathbb{I}_{N}\otimes T^{a} + \mathcal{C}_{ab}T^{a}\otimes T^{b},
\label{eq:M_decomp}
\end{equation}
where the $T^{a}$ are the $N^{2}-1$ generators of $SU(N)$ in the defining representation under the conventions outlined in Appendix~\ref{sec:appendix_A}. We will refer to this operator basis as the \textit{qudit} basis. The use of this basis does not impose any symmetry properties on the amplitudes, but rather results in the fact that the set $\{\mathbb{I},T^{a}\}$ form a linearly independent set of order $N^2$ on each factor of $H_{N}\otimes H_{N}$. The coefficients of this expansion are defined by
\begin{flalign}\nonumber
    \tilde{\mathcal{M}}=&\frac{1}{N^{2}}\text{Tr}(\mathcal{M}\mathbb{I}_{N}\otimes\mathbb{I}_{N}),\quad\quad\mathcal{A}_{a}=\frac{2}{N}\text{Tr}(\mathcal{M}T^{a}\otimes\mathbb{I}_{N})\\\
    \mathcal{B}_{a}=&\frac{2}{N}\text{Tr}(\mathcal{M}\mathbb{I}_{N}\otimes T^{a}), \quad\quad\frac{\mathcal{C}_{ab}}{4}=\text{Tr}(\mathcal{M}T^{a}\otimes T^{b}).
\end{flalign}

Expanding the amplitude in the qudit basis provides an easy way to isolate the leading effects of entanglement generation of a given scattering process. Without imposing any specific constraints on the amplitude, we may isolate various matrix elements which generate non-local correlations in the qudit space by
\begin{flalign}
\mathcal{M}_{kl,ij} =&\; \mathcal{C}_{ab}T^{a}_{ki}T^{b}_{lj}, \quad\quad\quad\quad\quad\quad\; i\neq k,\; j\neq l,\label{eq:C_coeffs1}\\
\mathcal{M}_{il,ij} - \mathcal{M}_{kl,kj} =& \;\mathcal{C}_{ab}(T^{a}_{ii}T^{b}_{lj} - T^{a}_{kk}T^{b}_{lj}),\quad\quad\quad\; j\neq l,\label{eq:C_coeffs2}\\
\mathcal{M}_{kj,ij} - \mathcal{M}_{kl,il} =& \;\mathcal{C}_{ab}(T^{a}_{ki}T^{b}_{jj} - T^{a}_{ki}T^{b}_{ll})\quad\quad\quad\;\; k\neq i,\label{eq:C_coeffs3}\\
\mathcal{M}_{ij,ij} + \mathcal{M}_{kl,kl} - \mathcal{M}_{il,il} - \mathcal{M}_{kj,kj} =& \;\mathcal{C}_{ab}(T^{a}_{ii}T^{b}_{jj} + T^{a}_{kk}T^{b}_{ll} - T^{a}_{ii}T^{b}_{ll} - T^{a}_{kk}T^{b}_{jj}).
\label{eq:C_coeffs4}
\end{flalign}
In the limit $\mathcal{C}_{ab}=0$, we recover the conditions found in~\cite{Carena:2023vjc,Chang:2024wrx} for the amplitude to preserve the set of separable states, up to higher order effects in their respective choices of entanglement measures. These relations have recently been recovered by independent calculation which generalized these observations to mixed final states in the qudit quantum numbers~\cite{Kowalska:2025qmf}. In the following, we will use $\mathcal{C}_{ab}=0$ as a useful diagnostic for the amplitude to preserve separability.

Consider now scattering amplitudes with an $SU(N)$ global symmetry in the qudit space. First let us consider particle-particle scattering on the $\mathbf{N}\otimes \mathbf{N}$ space. As we have argued, any $SU(N)$-invariant amplitude can be decomposed as
\begin{equation}
	\mathcal{M}=\mathcal{M}_{\textbf{S}}P_{\textbf{S}} + \mathcal{M}_{\textbf{A}}P_{\textbf{A}},
\end{equation}
where $P_{\textbf{S/A}}$ are again the projectors into the symmetric and antisymmetric invariant subspaces. Note that this two dimensional space of projectors can be recast into a reducible form
\begin{flalign}
	P_{\textbf{S}} + P_{\textbf{A}} =& \mathbf{S}_{\mathbb{I}},\\
	P_{\textbf{S}} - P_{\textbf{A}} =& \mathbf{S}_{W}
\end{flalign}
where $\mathbf{S}_{\mathbb{I}} = \mathbb{I}_{N}\otimes\mathbb{I}_{N}$, and we have introduced the swap gate
\begin{equation}
	\mathbf{S}_{W} = \sum_{i,j} \ket{i,j}\bra{\bar{j},\bar{i}}, \quad\quad     \mathbf{S}_{W} = \frac{1}{N}\left(\mathbb{I}_{N}\otimes\mathbb{I}_{N} + 2N\sum_{a=1}^{N^{2}-1}T^{a}\otimes T^{a}\right).
    \label{eq:swap}
\end{equation}

Although the projection operators are Hermitian and satisfy $P\cdot P = P$, their reducible forms are unitary
\begin{equation}
	(P_{\textbf{S}} \pm P_{\textbf{A}})^{\dagger}(P_{\textbf{S}} \pm P_{\textbf{A}}) = P_{\textbf{S}} + P_{\textbf{A}} = \mathbf{S}_{\mathbb{I}}.
\end{equation}
From this point of view the invariance of $\mathbf{S}_{W}$ naturally arises as a reducible representation of the diagonal action of $SU(N)\times SU(N)$ on $\mathbf{N}\otimes \mathbf{N}$ space. Thus, any $SU(N)$ invariant amplitude acting on this space can be put into the reducible form
\begin{equation}
	\mathcal{M} = \frac{1}{2}\left(\mathcal{M}_{\textbf{S}} + \mathcal{M}_{\textbf{A}}\right)\mathbf{S}_{\mathbb{I}} + \frac{1}{2}\left(\mathcal{M}_{\textbf{S}} - \mathcal{M}_{\textbf{A}}\right)\mathbf{S}_{W}.
\end{equation}

Computing the $\mathcal{C}_{ab}$, we find
\begin{equation}
	\mathcal{C}_{ab} = (\mathcal{M}_{\textbf{S}} - \mathcal{M}_{\textbf{A}})\delta_{ab}
\end{equation}
aligning with the arguments of~\cite{Carena:2023vjc} that the $\mathbf{S}_{W}$ operator itself will generically generate entanglement in the context of perturbative scatting amplitudes where $\mathbf{S}\sim \mathbb{I} + i\mathcal{M}$, and satisfying Eqs.~\ref{eq:C_coeffs1}-~\ref{eq:C_coeffs4} requires the amplitude to be in the equivalence class of the identity.~\footnote{Although we have expressed these relations in the computational basis, a change of basis to another element in the equivalence class does not change these conclusions. From the point of view of the qudit space this follows from the fact that $\ket{i}\otimes\ket{j}\simeq \ket{i}\otimes U\ket{j}$, for $U\in SU(N)$, and clearly $\mathbf{N}^{\prime} = U\ket{j}$ is still a fundamental representation. In this basis, we may expand $\mathbf{N}\otimes \mathbf{N}^\prime = \textbf{S}\otimes\textbf{A}$, and an $SU(N)$ invariant amplitude will act diagonally in this space. From the point of view of entanglement, this follows from the fact that the amount of entanglement generated, parameterized by $\mathcal{C}_{ab}$, is the same for every member of the equivalence class.}

Moving on to scattering on the $\mathbf{N}\otimes \bar{\mathbf{N}}$ space, the invariant amplitudes can be decomposed as
\begin{equation}
\mathcal{M}=\mathcal{M}_{\mathbf{1}}P_{\mathbf{1}} + \mathcal{M}_{\mathbf{Adj}}P_{\mathbf{Adj}},
\end{equation}
where $P_{\mathbf{1/Adj}}$ are now the projectors into the singlet and adjoint subspaces from pairing incoming and outgoing states. As in the previous case, we may recast this two dimensional space of projectors into their reducible form
\begin{flalign}
	P_{\textbf{1}} + P_{\textbf{Adj}} =& \;\mathbf{S}_{\mathbb{I}},\\
	P_{\textbf{1}} - P_{\textbf{Adj}} =& \;\mathbf{U},
\end{flalign}
where
\begin{equation}
	\mathbf{U} = \left(\frac{2}{N^{2}}-1\right)\mathbb{I}_{N}\otimes\mathbb{I}_{N} - \frac{4}{N}\sum_{a}T^{a}\otimes T^{a}.
\end{equation}
As before the reducible forms are unitary $(P_{\textbf{1}} \pm P_{\textbf{Adj}})^{\dagger}(P_{\textbf{1}} \pm P_{\textbf{Adj}}) = \mathbf{S}_{\mathbb{I}}$. Thus, writing the amplitude in its reducible form we have
\begin{equation}
	\mathcal{M} = \frac{1}{2}\left(\mathcal{M}_{\textbf{1}} + \mathcal{M}_{\textbf{Adj}}\right)\mathbf{S}_{\mathbb{I}}+ \frac{1}{2}\left(\mathcal{M}_{\textbf{1}} - \mathcal{M}_{\textbf{Adj}}\right)\mathbf{U}.
\end{equation}
and now
\begin{equation}
	\mathcal{C}_{ab} = \frac{4}{N}\delta_{ab}.
\end{equation}
We note that the action of $\mathbf{U}$ on the irreducible subspaces is quite simple
\begin{flalign}
	\mathbf{U}\ket{\psi}_{\textbf{1}}= +\ket{\psi}_{\textbf{1}},\quad\quad
	\mathbf{U}\ket{\psi^{a}}_{\textbf{Adj}} = - \ket{\psi^{a}}_{\textbf{Adj}}.
\end{flalign}
Thus, the action of $\mathbf{U}$ realizes a charge-parity between singlet and adjoint states in the $\mathbf{N}\otimes \bar{\mathbf{N}}$ space.

We may define a basis of states corresponding to the reducible form by
\begin{flalign}
	\ket{\alpha} =& \ket{\psi}_{\textbf{1}} + \sum_{a}\ket{\psi^{a}}_{\textbf{Adj}}\\
	\ket{\beta} =& \ket{\psi}_{\textbf{1}} - \sum_{a}\ket{\psi^{a}}_{\textbf{Adj}}
\end{flalign}
On this two-dimensional space, the operator $\mathbf{U}$ acts as an exchange
\begin{equation}
	\mathbf{U}\ket{\alpha} = \ket{\beta},\quad\quad \mathbf{U}\ket{\beta}=\ket{\alpha}
\end{equation}	
A similar Swap-like operator was previously found in the context of entanglement suppression of nuclear systems~\cite{Hu:2025lua}. Investigation of $\mathbf{U}$ as a quantum gate and corresponding algorithms will be presented elsewhere.

Let us pause to summarize these results. Any 2-2 scattering amplitude on the qudit space which possesses an $SU(N)$ global symmetry can be put into one the, albeit reducible, forms
\begin{flalign}
\mathcal{M} =& \mathcal{M}_{\mathbf{N}\otimes \mathbf{N}}\mathbf{S}_{\mathbb{I}}+ \mathcal{M}_{W}\mathbf{S}_{W},\\
\mathcal{M} =& \mathcal{M}_{\mathbf{N}\otimes \bar{\mathbf{N}}}\mathbf{S}_{\mathbb{I}} + \mathcal{M}_{U}\mathbf{U}.
\end{flalign}
Thus, the space of 2-2 scattering amplitudes is completely spanned by only 3 quantum gates, $\mathbf{S}_{\mathbb{I}},\mathbf{S}_{W},$ and $\mathbf{U}$, and entanglement is generated by the presence of $\mathbf{S}_{W}$ and $\mathbf{U}$. The discussion related to recoupling relations for $SU(N)$ scattering amplitudes imposes analogous relations for the quantum gates. We find that the quantum gates spanning $SU(N)$ scattering amplitudes obey the following recoupling relations
\begin{equation}
	\begin{pmatrix}\mathbf{I}\\ \mathbf{S}_{W} \end{pmatrix} = \begin{pmatrix}\frac{N}{2} &\frac{N}{2}\\ 1 & 0 \end{pmatrix}\cdot \begin{pmatrix}\mathbf{I} \\\mathbf{U} \end{pmatrix}.
	\label{eq:crossing_gates}
\end{equation}

Remarkably, realizing the action of $SU(N)$ invariant scattering amplitudes as quantum gates reveals that, in fact, any 2-2 scattering amplitude is a manifestation of the action of the cyclic group of order 2, $\mathbb{Z}_{2}$.~\footnote{Mathematically, one can view the appearance of $\mathbb{Z}_2$ on ${\mathbf{N}\otimes \mathbf{N}}$ from its isomorphism to the permutation group, $\mathcal{S}_{2}$. On $\mathbf{N}\otimes \bar{\mathbf{N}}$, the set $\{\mathbf{S}_{\mathbb{I}}, \mathbf{U}\}$ furnishes a representation of the outer automorphism group of the diagonal action of $SU(N)\times SU(N)$~\cite{FultonHarris:1991}. We will not elaborate on these details here, but simply rely on our information-theoretic perspective.} On the $\mathbf{N}\otimes \mathbf{N}$ ($\mathbf{N}\otimes \bar{\mathbf{N}}$) space this can be seen from the fact that $\mathbf{S}_{W}^{2}=\mathbf{S}_{\mathbb{I}}$ ($\mathbf{U}^{2}=\mathbf{S}_{\mathbb{I}}$). Further, crossing implies that these two expansions are not independent. We will use these relations to study the entanglement predicted by crossing symmetry across different 2-2 scattering channels.
\section{Entanglement from crossing symmetry}
\label{sec:4}
In this section, we provide our main conclusions that crossing symmetry requires the generation of entanglement in at least one scattering channel for 2-2 scattering of theories with an $SU(N)$ global symmetry. Let us start by considering scattering in the $s$-channel which we define to be the invariant amplitude acting on the $\mathbf{N}\otimes \mathbf{N}$ qudit space
\begin{flalign}\nonumber
\mathcal{M}^{(s)} =\; \mathcal{A}^{s}\mathbf{S}_{\mathbb{I}} + \mathcal{B}^{s}\mathbf{S}_{W},
\end{flalign}
Recall that in this form, the amount of entanglement generated by this amplitude is parameterized by
\begin{equation}
	\mathcal{C}_{ab} = 2\mathcal{B}^{s}\delta_{ab}
\end{equation}
Thus, requiring that the amplitude preserve the set of separable states up to the action of local unitaries requires $ \mathcal{B}^{s} = 0$. From the recoupling relations of quantum gates, Eq.~\ref{eq:crossing_gates}, plus crossing symmetry we find a prediction for the $t$-channel amplitude
\begin{flalign}
\mathcal{M}^{(t)} = \frac{\mathcal{A}^{t}}{2}(\mathbf{S}_{\mathbb{I}} + \mathbf{U})
\end{flalign}
where $\mathcal{A}^{t}$ is the analytically continued, crossed-recoupled amplitude of $\mathcal{A}^{s}$
Thus, in the $t$ channel
\begin{equation}
	\mathcal{C}^{(t)}_{ab} = 2\mathcal{A}^{t} \text{Tr}(\mathbf{U}T^{a}\otimes T^{b})\neq0
\end{equation}
 We see that demanding entanglement suppression in the $s$-channel guarantees that entanglement is generated in the $t$-channel by crossing symmetry. Since the analytically continued ampltidue $\mathcal{A}^{t}$ is fundamentally the same complex function as $\mathcal{A}^{s}$ evaluated in different regions of the physical on-shell kinematics, further demanding entanglement suppression in the $t$-channel leads to $\mathcal{M}=0$.
 
 If we had instead started by imposing entanglement suppression in the $t$-channel
 \begin{equation}
	 \mathcal{M}^{(t)} = \mathcal{A}^{t}\mathbf{S}_{\mathbb{I}} 
 \end{equation}
 Then, applying the same manipulations of crossing to the $s$-channel we find
 \begin{equation}
 	\mathcal{M}^{(s)} = \mathcal{A}^{s}\mathbf{S}_{W}.
 \end{equation}
As we have noted, although $\mathbf{S}_{W}$ itself preserves the set of separable states up to the equivalence class, the presence of this operator in the $S$-matrix will generically generate some amount of entanglement. We can see this schematically, for instance, by starting in the pure qudit  state $\ket{01}$. The final state after action of the operator $\mathbf{S}_{\mathbb{I}}+ \mathcal{A}\mathbf{S}_{W}$
 \begin{equation}
 \ket{f}=\ket{01} +  \mathcal{A}\ket{10}
 \end{equation}
 which is clearly not separable and will have some entanglement for $\mathcal{A} \neq 0$.

These constraints follow independently of other details of  a specific model, e.g. Bose symmetry or Fermi-Dirac statistics, or the inclusion of gauge charges. In specific models these additional details can lift requirements related to identical or non-identical particle scattering etc. Nevertheless, the amplitudes in the qudit quantum numbers must satisfy the crossing relations we have just derived if there is an underlying $SU(N)$ global symmetry. Although in some cases additional details of the model might automatically force the relations to be satisfied trivially. Our results also extend to cases such as $(\textit{scalar})_{i} + (\textit{fermion})_{j} \to (\textit{scalar})_{k} + (\textit{fermion})_{l}$ scattering, or scattering of any combination of helicities for that matter, so long as the qudit numbers $i,j,k,l = 1,..,N$ transform in the fundamental or antifundmental representations.~\footnote{These conclusions likely extend to other measures of quantum correlations. For instance, ref~\cite{Busoni:2025dns} proposed that amplitudes having the form $\mathcal{M}\propto \mathbf{S}_{\mathbb{I}}$ conserve quantum magic, finding that this constraint is correlated with an enhanced symmetry in a 2HDM. As we have seen, amplitudes of the form $\mathcal{M}\propto \mathbf{S}_{\mathbb{I}}$ predict other tensor projections in the crossed channels. We leave investigation of crossing to other quantum correlations to future work.} 

Any interacting theory which possesses an $SU(N)$ global symmetry and nontrivial interactions generates entanglement in at least one scattering channel in the bipartite space of qudits.
\section{Discussion and Conclusions} 
\label{sec:conclusions}

In this paper, we establish a concrete bridge between two foundational principles of quantum field theory and quantum information: crossing symmetry and entanglement. By treating scattering amplitudes as quantum operations acting on the Hilbert space of internal quantum numbers, we have shown that the entangling structure of the $S$-matrix is intimately connected to the same analytic properties that underlie locality and causality. In this framework, crossing symmetry enforces relations between scattering channels that forbids the complete suppression of entanglement. Only free, non-interacting theories avoids this interplay. If entanglement suppression is part of an information-theoretic principle from which symmetry is an emergent property, it must be a channel specific phenomenon.

Recasting $SU(N)$-invariant amplitudes as compositions of quantum gates exposes a remarkably simple algebraic structure. All $2\to2$ scattering processes invariant under $SU(N)$ are generated by a minimal set of three unitary operations, $\{\mathbf{S}_{\mathbb{I}}, \mathbf{S}_{W}, \mathbf{U}\}$, which on each scattering channel realize a $\mathbb{Z}_{2}$ action. The unitary, albeit reducible, realization of this action for particle-particle scattering, where $\mathbb{Z}\to \mathbf{S}_{W}$, the action of this group manifests as spin-statistics, while for particle-antiparticle scattering, where $\mathbb{Z}\to \mathbf{U}$, it manifests as a charge-conjugation parity. 

These observations point to deeper organizing principles at the intersection of quantum information and field theory: symmetry and entanglement $S$-matrix are fundamentally interchangeable aspects of its analytic and algebraic structure. The emergence of $SU(N)$ symmetry can be seen as the condition that the entangling operations of the $S$-matrix close under the $\mathbb{Z}_{2}$ group, while crossing symmetry ensures their universality across channels. Together they hint at a new information-theoretic principle governing the structure of consistent quantum field theories, one in which the generation, suppression, and transformation of entanglement are intrinsically related to analytic aspects of the $S$-matrix.

Although our treatment has been rigorous, there are some limitations. Principally, we have implicitly relied on the proof of crossing symmetry for $2\to2$ scattering amplitudes. From the work of Bros, Epstein, and Glaser, this implies that these results strictly hold for theories with a mass gap. Their results also extend to $2\to 3$ amplitudes~\cite{Bros:1985gy}, and so a natural continuation of this work would lie in extending this framework to higher-point amplitudes. Further, even for $2\to 2$ scattering there is work to be done in exploring scattering between Hilbert spaces of differing dimensions $H_{N}\otimes H_{M}$, or considering embedding of subgroups into the Hilbert space, e.g. $\mathbf{2}\otimes\mathbf{3}\sim\mathbf{5}$, for $\mathbf{N}=5$ in $SU(5)$. Finally, while the present work has leaned toward implications of crossing for entanglement suppression, it is of interest to extend these considerations to other information-theoretic principles related to maximal entanglement, or quantum magic. We plan to revisit these questions in future work.

More broadly, these findings suggest that the analytic structure of quantum field theory may be imprinted in information-theoretic principles. In particular crossing symmetry, regarded as a heuristic physical principle requiring particles traveling forward in time to be indistinguishable from antiparticles traveling backward in time, is deeply intertwined with the generation of entanglement in scattering processes that possess an $SU(N)$ symmetry.

\vspace{0.05cm}

\acknowledgments
I thank Andrew Jackura for comments on an earlier draft, and Ian Low for numerous discussions.
This work has been supported in part by the U.S. Department of Energy under grant No. DEFG02-13ER41976/DE-SC0009913.

\appendix

\section{Projection operators and recoupling relation for SU(N)}
\label{sec:appendix_A}
\subsection{Two qudit states}
\label{sec:sub_app_A1}
In this appendix, we review definitions of projection operators and recoupling relations relevant for $SU(N)$ irreducible tensor decompositions. We assume that the qudit states, $\ket{i}$, transform as irreducible representations of $SU(N)$. The generators of the Lie algebra of $SU(N)$ in the computational basis are 
 \begin{flalign}\nonumber
    T^k &= \frac{1}{\sqrt{2k(k+1})}\left(\sum_{i=1}^{k}\ket{i}\bra{i} -k\ket{k+1}\bra{k+1}\right),\\\nonumber
    &T^{+}(i;j) =\frac{1}{2}\left(\ket{i}\bra{j}+\ket{j}\bra{i}\right),\\
    &T^{-}(i;j) =\frac{i}{2}\left(\ket{i}\bra{j}-\ket{j}\bra{i}\right).
\end{flalign}
which satisfy the brackets $[T^{a},T^{b}]=if_{abc}T^{c}$. The normalization has been chosen so that $\text{Tr}(T^{a}T^{b})=\frac{\delta_{ab}}{2}$~\cite{Georgi:1982jb,Ramond:2010zz}. \\

Generically, a two-qudit state, $\ket{i}\otimes\ket{j}$, can be decomposed into a direct sum of irreducible representations
\begin{equation}
\ket{i}\otimes\ket{j}\simeq \bigoplus_{R}R
\label{eq:irrep_decomp}
\end{equation}
where $R$ labels the irreducible representation. The two-qudit states in the computational basis can be expressed as

\begin{equation}
	\ket{ij} = \sum_{R,r}C^{ij}_{R,r}\ket{R,r}  = \sum_{\substack{R,r\\k,l}}C^{ij}_{R,r}\left(C^{kl}_{R,r}\right)^{*}\ket{kl},
	\label{eq:clebsch}
\end{equation}
where $R$ sums over all irreducible representations in the decomposition Eq.~\ref{eq:irrep_decomp}, and $r=1,...,\text{dim}(R)$. In the second step, we have inserted a complete set of two-qudit states. This defines the relations for the projection operators for the irreducible representation $R$ as 
\begin{equation}
	P_{R}^{ij,kl}\equiv \sum_{r} C^{ij}_{R,r}\left(C^{kl}_{R,r}\right)^{*},\quad\quad\quad \sum_{R} P_{R} = \mathbb{I}
\end{equation}
Note that depending on the representations of the states $\ket{i}$, some irreducible representations, $R$, may appear with multiplicity $N_{R}$ in the sum Eq.~\ref{eq:clebsch}. In this case, an additional indexing over the multiplicities should be included $\sum_{R,r}C^{ij}_{R,r}\to \sum_{R,n,r}C^{ij}_{R_n,r}$, where $n=1,..., N_{R}$, etc. For the case of two-qudit states in the fundamental representation, the decomposition into irreps under $SU(N)$ is
\begin{equation}
	N\otimes N = \mathbf{S}\oplus\mathbf{A}.
\end{equation}
The symmetric and antisymmetric states are defined by
\begin{equation}
\ket{\psi^{ij}}_{S} = \frac{1}{2}\left(\ket{ij} + \ket{ji}\right),\quad\quad\quad\quad \ket{\psi^{ij}}_{A} = \frac{1}{2}\left(\ket{ij} - \ket{ji}\right),
\end{equation}
with corresponding Clebsch-Gordon coefficients given by
\begin{equation}
	C^{ij}_{\textbf{S},kl} = \frac{1}{2}\left(\delta_{ki}\delta_{lj} + \delta_{li}\delta_{kj}\right)\quad\quad\quad\quad 	C^{ij}_{\textbf{A},kl} = \frac{1}{2}\left(\delta_{ki}\delta_{lj} - \delta_{li}\delta_{kj}\right)
\end{equation}

For a two qudit state combining a fundamental and anti-fundamental rep we have
\begin{equation}
	N\otimes\bar{N} = \mathbf{1}\oplus\mathbf{Adj}.
\end{equation}
The orthonormal basis of singlet and adjoint states are then defined by
\begin{equation}
	\ket{\psi}_{\textbf{1}} = \frac{1}{\sqrt{N}}\sum_{k}\ket{k\bar{k}}, \quad\quad\quad\quad \ket{\psi^{a}}_{\textbf{Adj}} = \sqrt{2} \sum_{ij}T^{a}_{kl}\ket{k\bar{l}},
\end{equation}
With Clebsch-Gordon coefficients given by
\begin{equation}
 	C^{ij}_{\textbf{1}} = \;_\textbf{1}\braket{\psi|ij} = \frac{\delta_{ij}}{\sqrt{N}}, \quad\quad\quad\quad  	C^{ij}_{\textbf{Adj},a} = \;_\textbf{Adj}\braket{\psi^{a}|ij} = \sqrt{2}T^{a}_{ij}
\end{equation}
Since these two cases are our main focus and all the necessary irreps appear only once in the decompositions, we will suppress the multiplicity label in the following.

\subsection{Projection operators and recoupling relations}
\label{sec:sub_app_A2}
Consider a bipartite operator $A$ acting on $H_{N}\otimes H_{\bar{N}}$ with matrix elements
\begin{flalign}
	\bra{\bar{k}l}A\ket{i\bar{j}}&=A_{\bar{k}li\bar{j}}\\
					       & =  \left[P_{\textbf{S}}+ P_{\textbf{A}}\right]_{li,pq}\left[P_{\textbf{S}}+ P_{\textbf{A}}\right]_{kj,rs}A_{rp,qs}.
\end{flalign}
If $A$ is an invariant tensor, then the cross terms $\sim P_{\textbf{S}}P_{\textbf{A}}$ vanish
\begin{equation}
	\bra{\bar{k}l}A\ket{i\bar{j}} = \left[\left(P_{\mathbf{S}}\right)_{li,pq} \left(P_{\mathbf{S}}\right)_{kj,rs} + \left(P_{\mathbf{A}}\right)_{li,pq}\left(P_{\mathbf{A}}\right)_{kj,rs}\right]A_{rp,qs}.
\end{equation}
Now, let $A$ be a projection operator in the $s$ channel, $P^{(s)}_{\mathbf{1,Adj}}$. Then we obtain
\begin{flalign}
\bra{\bar{k}l}P^{(s)}_{\mathbf{1}}\ket{i\bar{j}} &=\left[\left(P_{\mathbf{S}}\right)_{li,pq} \left(P_{\mathbf{S}}\right)_{kj,rs} + \left(P_{\mathbf{A}}\right)_{li,pq}\left(P_{\mathbf{A}}\right)_{kj,rs}\right]\frac{\delta_{pr}\delta_{qs}}{N}\\
							& = \frac{1}{2N}\left[\left(\delta_{lk}\delta_{ij} + \delta_{ik}\delta_{lj}\right) + \left(\delta_{lk}\delta_{ij} - \delta_{ik}\delta_{lj}\right)\right]\\
							&=\frac{1}{N}(P^{(t)}_{\mathbf{S}})_{li,kj} + \frac{1}{N}(P^{(t)}_{\mathbf{A}})_{li,kj}.
\end{flalign}
Similarly for the adjoint projector we find
\begin{equation}
	\bra{\bar{k}l}P^{(s)}_{\mathbf{Adj}}\ket{i\bar{j}} = \frac{N-1}{N}(P^{(t)}_{\mathbf{S}})_{li,kj} - \frac{N+1}{N}(P^{(t)}_{\mathbf{A}})_{li,kj}
\end{equation}

\begin{equation}
	\begin{pmatrix}P^{(s)}_{\mathbf{1}} \\P^{(s)}_{\mathbf{Adj}} \end{pmatrix} = \mathbf{C}_{s,t}\cdot \begin{pmatrix}P^{(t)}_{\mathbf{S}} \\P^{(t)}_{\mathbf{A}} \end{pmatrix},
\end{equation}
where we have introduced the crossing matrix
\begin{equation}
\renewcommand\arraystretch{1.8}
	\mathbf{C}_{s,t}=\begin{pmatrix}\frac{1}{N} &\frac{1}{N}\\ \frac{N-1}{N} & -\frac{N+1}{N} \end{pmatrix}
\end{equation}
Similarly
\begin{equation}
\renewcommand\arraystretch{1.8}
	\mathbf{C}_{t,s}=\begin{pmatrix}\frac{N+1}{2} &\frac{1}{2}\\ \frac{N-1}{2} & -\frac{1}{2} \end{pmatrix}
\end{equation}
Note
\begin{equation}
	\mathbf{C}_{s,t}\mathbf{C}_{t,s} =\begin{pmatrix}1&0\\0&1\end{pmatrix}
\end{equation}





\end{document}